\documentclass[pdflatex,sn-nature]{sn-jnl}

\usepackage[version=3]{mhchem} 
\usepackage{xcolor}
\usepackage{soul} 
\usepackage{array}
\usepackage{graphicx}%
\usepackage{multirow}%
\usepackage{amsmath,amssymb,amsfonts}%
\usepackage{amsthm}%
\usepackage{mathrsfs}%
\usepackage[title]{appendix}%
\usepackage{xcolor}%
\usepackage{textcomp}%
\usepackage{manyfoot}%
\usepackage{booktabs}%
\usepackage{algorithm}%
\usepackage{algorithmicx}%
\usepackage{algpseudocode}%
\usepackage{listings}%

\begin{document}

\title{Learning Between the Lines: Interaction Interpolation in a Multi Element Machine Learning Interatomic Potential}

\author{Ademola Soyemi$^{1,2}$}
\author{Kipton Barros$^{1,2}$}
\author{Timothy C. Germann$^1$}
\author{Benjamin T. Nebgen$^1$}
\author{Brenden W. Hamilton$^1$}
\affil{$^1$Theoretical Division, Los Alamos National Laboratory, Los Alamos, New Mexico 87545, USA}
\affil{$^2$Center for Nonlinear Studies, Los Alamos National Laboratory, Los Alamos, New Mexico 87545, USA}
\email{brenden@lanl.gov}
\date{June 2024}

\maketitle

\begin{abstract}

The continuous development of machine learning inter-atomic potentials (MLIAPs) has greatly accelerated materials research in recent years.
However, the need to construct custom MLIAPs for each unique problem remains a bottleneck which raises the question of the transferability of MLIAPs. 
While foundation MLIAPs can achieve transferability due to the broad coverage of chemical space represented in their training sets, their size and computational cost limits their routine use. 
We address this gap by developing a lightweight general-purpose MLIAP for metal alloys spanning twelve elements (Al, Sc, Ti, V, Cu, Zn, Zr, Nb, Mo, Hf, Ta, W). 
Here, we utilize a step-wise active learning (AL) approach to develop a compact yet diverse dataset and model that generalizes well by interpolating from previously seen interactions. 
The resulting potentials transfer effectively to unseen, higher body-order alloys systems, requiring only unary and binary training data to capture ternary and quaternary systems.
The model also transfers well to new elements via an alchemical extrapolation within the network, requiring only unary data for the new element and its alloys.
These findings suggest that MLIAPs trained on targeted, diversity-driven datasets provide transferability at a relatively low computational cost, offering a practical path toward general-purpose, resource-efficient MLIAPs for materials modeling.

\end{abstract}
\newpage

\section{Introduction}

Over the previous decade, machine learning (ML) has made great advances in accelerating scientific simulations,
especially through serving as surrogates for expensive computations \cite{rupp2012fast,desai2021parsimonious,butler2018machine,ramprasad2017machine,hamilton2023using,sahrmann2026bridging}.
Within materials modeling, 
particular attention has been paid to the development of machine learned inter-atomic potentials (MLIAPs) which are able to provide quantum mechanical accuracy at similar computational cost to classical molecular dynamics
\cite{smith2019approaching,zuo2020performance,smith2017ani,chmiela2018towards,lindsey2025chimes,f2025improving,sahrmann2026learning}. However, this assumes that the underlying training data is generated using a sufficiently high-level {\it ab initio} method,
which can lead to expensive start up costs \cite{schutt2017quantum,bartok2010gaussian,keith2021combining}.
Their accuracy in reproducing energies, forces, and stresses, as well as producing accurate dynamics, however, depends on not only the underlying quantum mechanical method but also on the training set composition.
\cite{kulichenko2024data,Maxson2024Practices,Maxson2025Materials25,schutt2017quantum,bartok2010gaussian,keith2021combining, Kim_Leveraging2026}. 

Among recent MLIAP research directions, two stand out: the automated curation of training data through active learning (AL) \cite{smith2018less,smith2021automated,kulichenko2024data,Yoo2021NNRF,hamilton2023high,vandermause2020fly,xie2021bayesian,montes2022training,hamilton2026stoichiometry}, 
and the push towards foundation or universal potentials whose datasets cover broad sections of the periodic table. \cite{batatia2025foundation,kovacs2025mace,allen2024learning,zhang2024exploring,huan2017universal,niblett2025transferability}. 
However, a major drawback of these foundation potentials is that they require large amounts of training data, incurring a high upfront computational cost. 
Additionally, foundation potentials typically contain millions of parameters which slows down inference, making routine molecular dynamics (MD) simulations considerably more computationally demanding. 
While targeted MLIAPs are generally preferred for studying specific systems, recent work has also shown that by careful curation, sufficient breadth can be achieved with surprisingly small datasets \cite{Maxson2025AgNP,Soyemi2025Na2So4,smith2021automated,zhang2024exploring}. 

Despite the efficiency of lightweight bespoke MLIAPs, a major open question is the transferability of these MLIAPs and the need to construct a new MLIAP for each unique problem. 
While foundation potentials accomplish this transferability through the sheer amount and breadth of their training data, a similar broad coverage of chemical space can be achieved by densely sampling key physical properties \cite{Soyemi2025Interface,smith2021automated,Maxson2025AgNP,zhang2024exploring} such that the MLIAP is able to interpolate unseen configurations. 

Previous attempts to create general-purpose MLIAPs for several elements, or even the entire periodic table have had mixed results \cite{song2024general}. 
While these potentials have been successively demonstrated for systems containing one or two species \cite{Bartok2018_Si,Rowe_Deringer2020_Carbon,Ugwumadu2023}, truly universal MLIAPs have only recently begun to be broadly utilized for specific applications \cite{Dicks2026,Sivak2025,Tateishi_PFP_2026} but are typically significantly slower than custom MLIAPs and classical force fields \cite{Takamoto2022,Chen_Ong2022}. 

Thus, in this work we develop a foundation-like, general-purpose MLIAP for metal alloys with a focus on reaching generality with a minimal amount of data. 
We show that a surprisingly small yet comprehensive dataset can enable a model to generalize well to new configurations through interpolating from already seen interactions.
Here, we apply the Active Learning Framework (ALF) code \cite{Grizzi_LohrALF2026} and conduct several stages of active learning (AL)
in which the nature of the available structures to be sampled and added to the training set are selectively controlled in each AL stage to optimize model learning, as opposed to brute force active learning whereby we sample from all available structures until model convergence or stability. 
Here, we show how initial models improve as data is added across the AL stages and demonstrate that higher body order data provides diminishing returns to the model. 
Alchemical extrapolation of multi-element interactions allows for the direct learning of ternary alloys from unary and binary data alone.
Additionally ,we show the transferability of these final models to quaternary systems, and that the training can be easily expanded to new elements and their alloys with minimal data. 

\section{Results}

Our training set generation is performed using an active learning-based, query-by-committee approach,
as has been previously utilized \cite{Grizzi_LohrALF2026}. 
As shown in Figure 1, there are four main steps in the active learning cycle: DFT calculations (labeling), Training Set Construction, Ensemble Training, and Uncertainty-based Dynamics (sampling).

\begin{figure}[htpb]
  \includegraphics[width=0.95\textwidth]{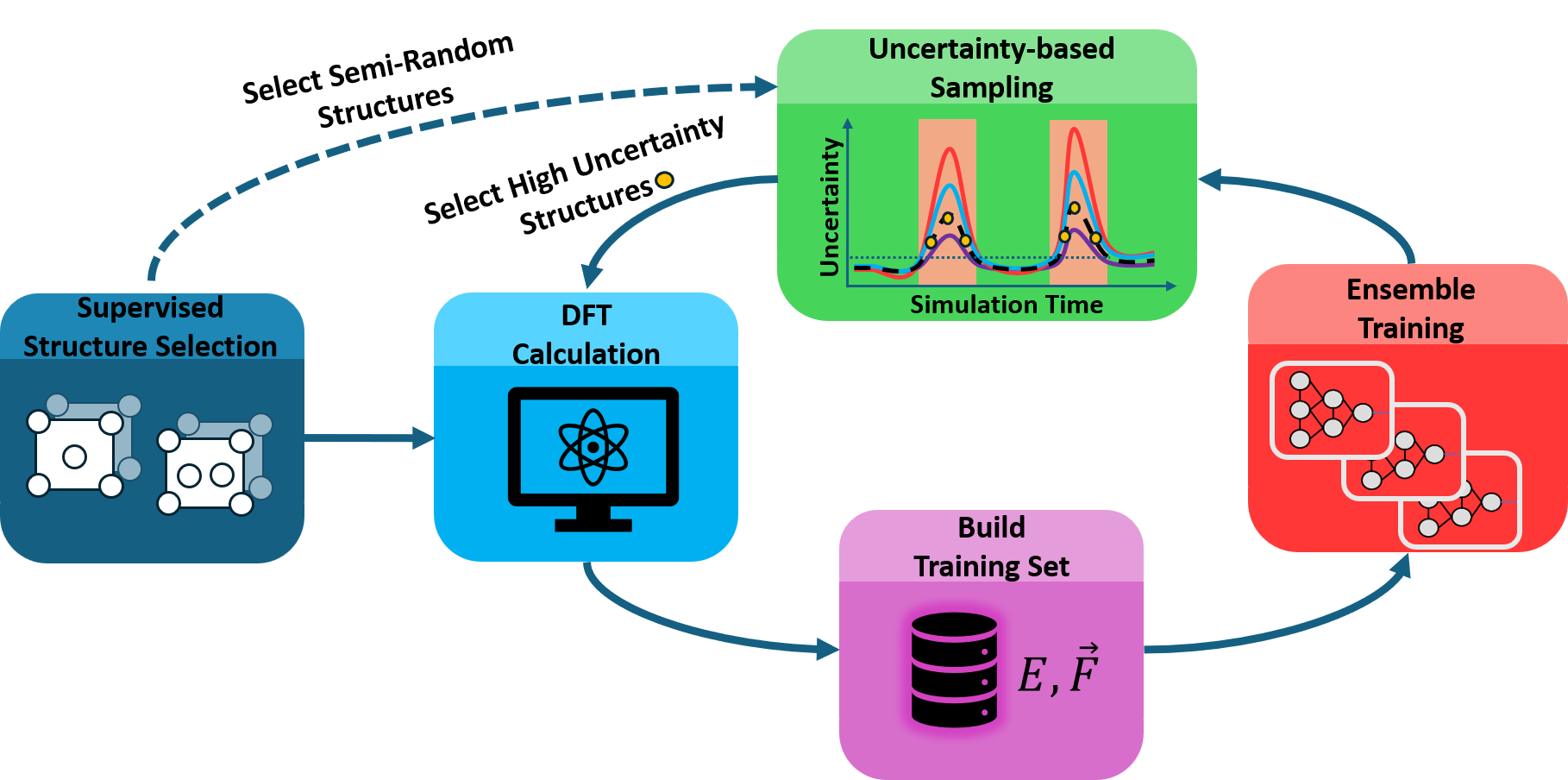}
  \caption{ALF-based active learning loop employed in this work.}
  \label{fig:Fg1}
\end{figure}

Our approach extends the existing query-by-committee active learning strategy {\cite{smith2018less,smith2019approaching,smith2021automated,Grizzi_LohrALF2026}} with a targeted bias that incorporates domain expertise to drastically enhances model learning at fixed data size budget. 
Here, we evolve the active learning (AL) structure selection routine over three stages: Unary only, Binary only, and Ternary only. 
It should be noted that data collected during each AL stage are retained and incorporated into model training in subsequent AL stages, but data is only appended to the data set from the current stage.
Each stage is further separated into two sub-stages: liquids and crystals.
Starting with the liquid has shown to dramatically improve data costs to run stable dynamics during AL, thereby accelerating discovery \cite{smith2021automated}. 
The transition from liquid to crystalline data occurs roughly halfway through each AL stage. 
Then sampling occurs on various crystal phases, volumes, and compositions, allowing us to sample the energy landscape and mechanical properties.
Our final data set includes only 10,131 total structures, including unary, binary, and ternary 
systems for 12 different metal elements: Al, Sc, Ti, V, Cu, Zn, Zr, Nb, Mo, Hf, Ta, W.
Binary and ternary alloys have their compositions varied over a number of ratios.
In all cases, the volume is varied by fractional re-mapping of atomic coordinates. 
See Supporting Information for additional details on the data generation procedure, and the structures and compositions used at each stage.


\subsection{Binary vs Ternary Training}

To evaluate the benefit of the ternary data on the performance of the model, we also launched a second AL run that stays in the binary phase such that we have a Unary-Binary (Figure 2, top) trained model and Unary-Binary-Ternary trained model (Figure 2, bottom) that have roughly the same amount of data and number of active learning cycles. 
Figure 3 shows the energy error of these two models along the AL routine, as the training sets grow, testing on all of the unary (blue), binary (red), or ternary (green) training data. 
Here, we train on exclusively unary data for 10 AL cycles, after which we introduce binary data (indicated by the vertical red dashed line in both panels of Figure 3) for 15 AL cycles. We highlight that at this point, Figure 3a and 3b are identical, however, in Figure 2a we continue to incorporate binary-only data for 25 additional AL cycles while in Figure 3b we introduce ternary-only data (indicated by the vertical green dashed line) for 25 AL cycles.

\begin{figure}[htpb]
  \includegraphics[width=0.95\textwidth]{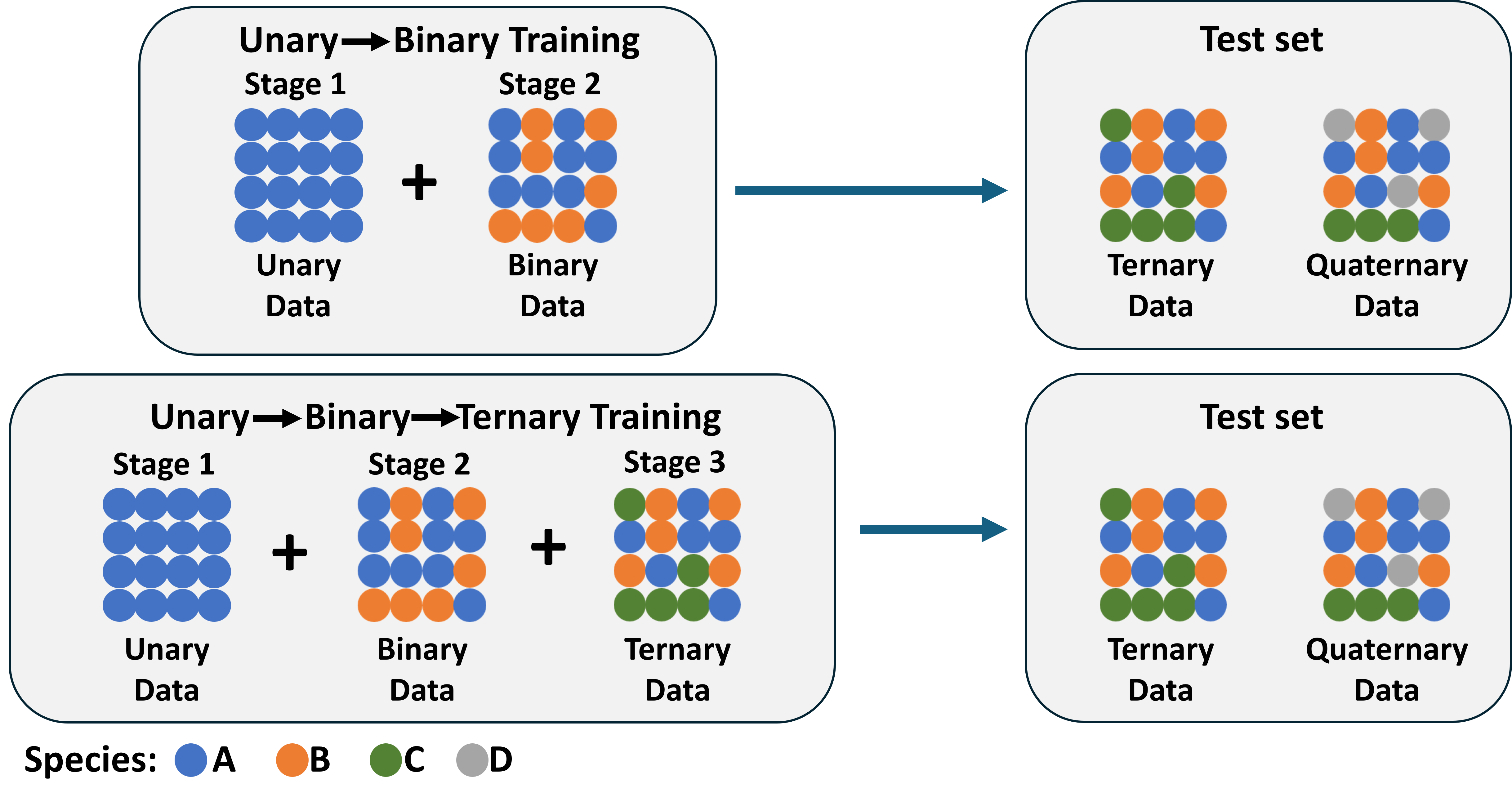}
  \caption{Active learning-based data accumulation and evaluation for Unary-Binary model (top panel), and Unary-Binary-Ternary model (bottom panel).}
  \label{fig:Fg2}
\end{figure}

Here, we observe that, in general, as we add more unary data, the energy errors progressively drop across all datasets. 
This shows that the model effectively learns simple interactions from the unary data which are transferable to the binary, and ternary data. 
After the full AL routine, the Unary-Binary and Unary-Binary-Ternary models (Figures 3a and 3b, respectively) yield the lowest energy errors for ternary data (0.33 and 0.30 eV/atom), with similar errors for unary (0.56 and 0.57 eV/atom) and binary (0.67 and 0.68 eV/atom) data.
For an ensemble of trained models, the difference in the Unary-Binary and Unary-Binary-Ternary models are within the model to model uncertainty. However, it is also important to note that the model errors in Figure 3b improve only very minimally upon addition of ternary data compared to Figure 3a, showing that the ternary data provide only marginal benefit and the model is already able to infer ternary configurations before seeing any ternary data. We attribute this behavior to unary and binary structures, containing 1- and 2- element interactions, spanning the relevant descriptor space with learned element embeddings potentially enabling alchemical generalization \cite{Csanyi2025} via interpolation to ternary combinations absent from the training data.

Interestingly, in both cases and for all stages of AL, we see the ternary error lowers in lock-step with the binary error, further supporting the inference that the model is inherently learning an alchemical extrapolation to higher body order interactions, and not just memorizing a force field. Ultimately, after the full AL cycle in both cases, the ternary errors are the lowest.
This is likely due to the ternary crystal data consisting only of structures with the BCC structure while the unary and binary data consist of structures with BCC, FCC, and HCP crystal symmetry in addition to the unary and binary data having a wider range of stoichiometric rations (see Table S1).

\begin{figure}[htpb]
  \includegraphics[width=0.95\textwidth]{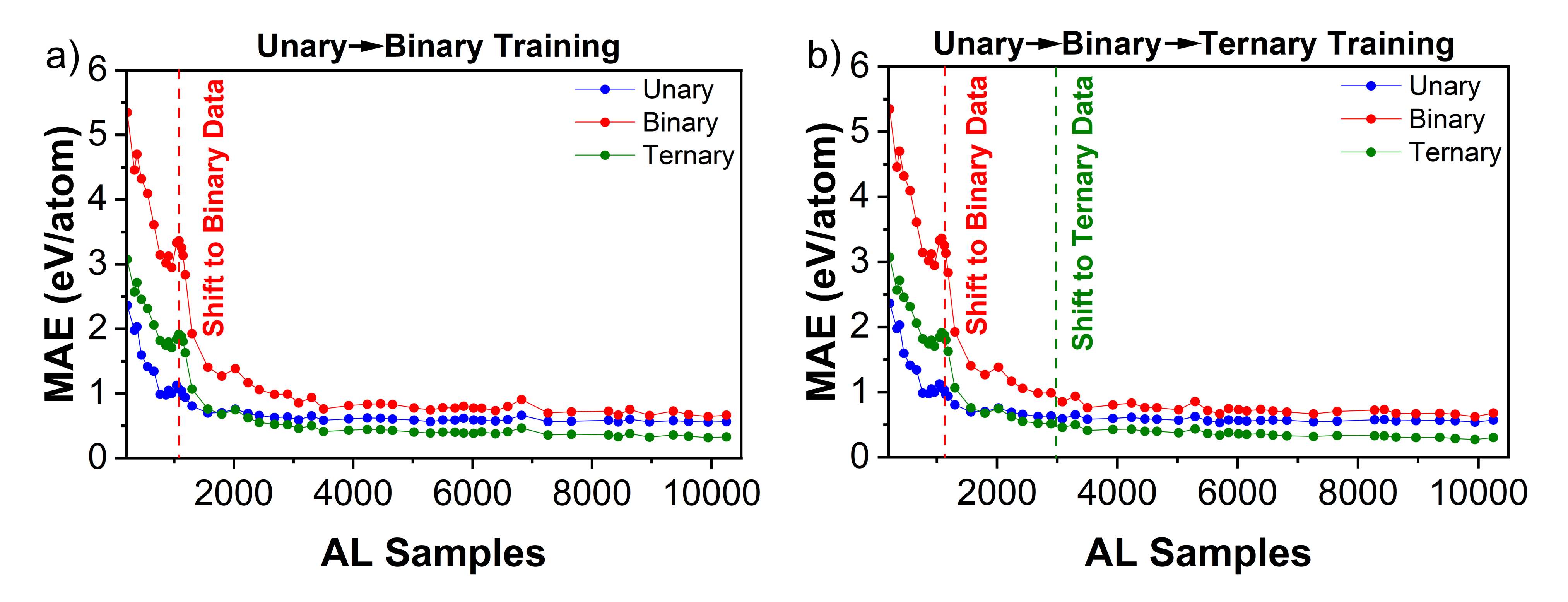}
  \caption{Prediction errors on the unary (blue), binary (red), and ternary (green) training data, respectively, as a function of total training data for the a) Unary-Binary, and b) Unary-Binary-Ternary models. Each point represents the a model trained during AL. Note that the energy errors in a) and b) are identical up till the point where ternary data is introduced (i.e., AL step 25 or at about 3000 data points, as depicted by the vertical green dashed line).}
  \label{fig:Fg3}
\end{figure}

Figure 4 below shows the test set errors for the final, hyperparameter-optimized Unary-Binary, and Unary-Binary-Ternary models. 
Here, the Unary-Binary model gave an energy MAE and RMSE of 8.55 meV/atom and 15.36 meV/atom, respectively, 
while the Unary-Binary-Ternary model gave an noticeably higher energy MAE and RMSE of 23.89 meV/atom and 61.59 meV/atom, respectively. 
While the Unary-Binary model showed lower errors, we note that its task and test set may be fundamentally easier than the Unary-Binary-Ternary model as the Unary-Binary model was trained and tested mainly on binary structures, as well as, the Unary-Binary-Ternary model errors are more outlier dominated.

\begin{figure}[htpb]
  \includegraphics[width=0.95\textwidth]{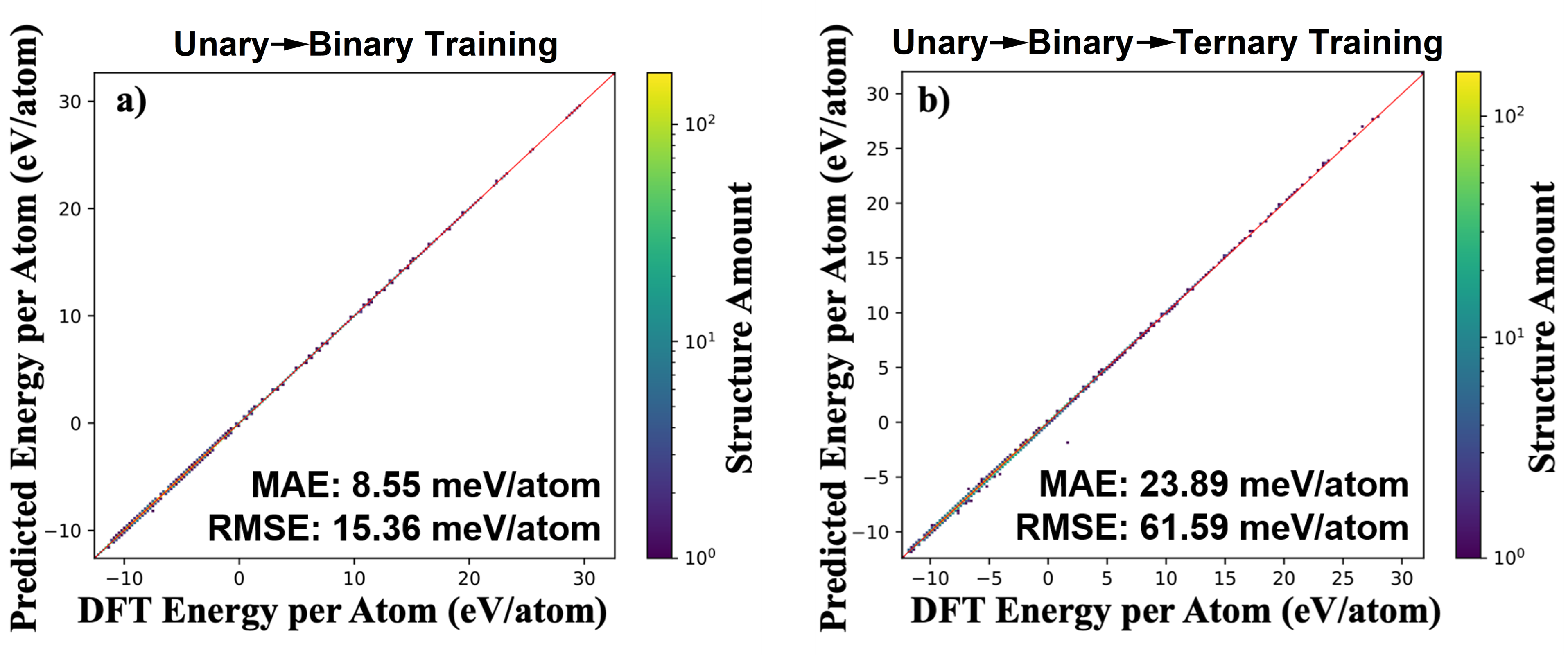}
  \caption{Parity plots comparing DFT and model predicted energy per atom for the final, hyper-parameter optimized a) Unary-Binary, and b)Unary-Binary-Ternary models on the test set.}
  \label{fig:Fg4}
\end{figure}

Lastly, we compared the Unary-Binary model's predictions to the Unary-Binary-Ternary model's predictions for all of the ternary structures, as shown in Figure 5 below.
In general, both models are congruent in their predictions shown in the R$^2$ value of 0.9955 and energy MAE of 0.06 meV/atom, and generally reproduce trends in the underlying DFT data (see Figure S2), showing that both models inherently "learn" the same things about the ternary structures. 
However, we observe several outliers that result in an energy RMSE of 0.23 eV/atom, with manual inspection of the worst outliers revealing that these structures predominantly consisted of vanadium-containing ternary alloys.
It should be noted that these are comparisons to each other, not to the ground truth, such that the divergence in any single prediction could be due to error in either model. 
We also find that the models diverge significantly in their predictions in the high energy range only shown in Figure S3. 

\begin{figure}[htpb]
  \includegraphics[width=0.5\textwidth]{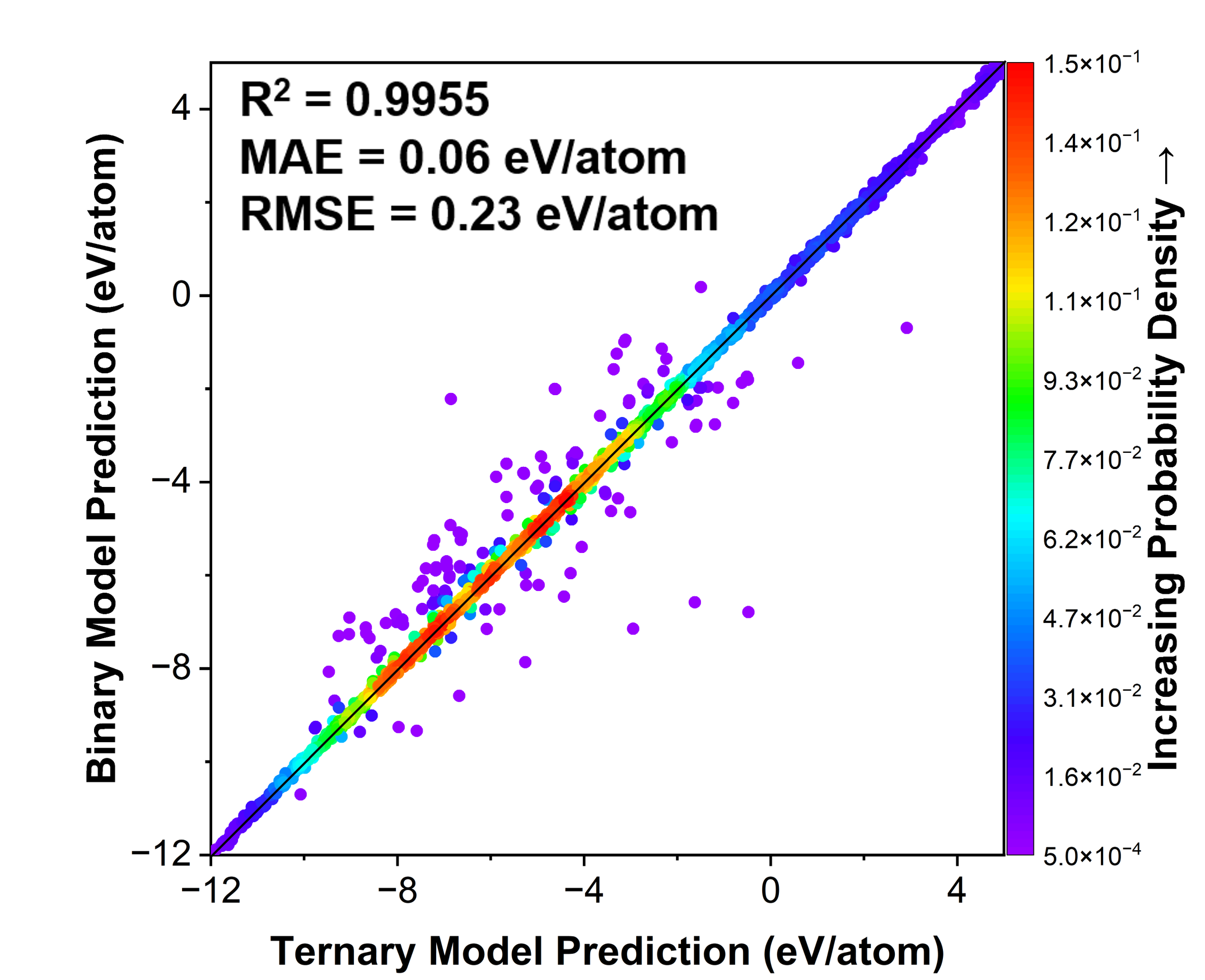}
  \caption{Parity plot comparing final hyper-parameter optimized Unary-Binary model and Unary-Binary-Ternary model predicted energies on ternary structures. Note that extremely high energy structures are not shown. See Figure S1 for full plot.}
  \label{fig:Fg5}
\end{figure}

\subsection{Extending to Four-Element Systems}

In the previous section, we showed that the Unary-Binary model is able to interpolate three-element interactions from unary and binary data only and give comparable performance to the Unary-Binary-Ternary model. 
To further assess how our models are learning alchemical extrapolations to higher body order interactions, we test our models on quaternary systems. 
Here, the final hyperparameter tuned models (Unary-Binary and Unary-Binary-Ternary) are tested on equi-atomic quaternary systems with the BCC crystal structure at a wide range of volumes for all combinations of the 12 metal elements. 
This quaternary data was generated using the same active learning routine used to generate the unary, binary, and ternary data whereby we scanned a wide range of volumes and random spatial distributions of elements, resulting in a dataset containing 1,582 structures. 

Figure 6 compares the model predictions to the DFT ground truth for low energy structures in the quaternary test set (see Figure S4 for full plot, in which extreme pressures at the used volume range leads to extreme positive energies). 
In general, we observe good predictive accuracy shown in the energy MAE of 0.06 eV/atom for both models. 
However, the Unary-Binary model predicts several outliers, resulting in a higher RMSE (0.24 eV/atom in Figure 6a) compared to the Unary-Binary-Ternary model (0.13 eV/atom in Figure 6b). Similar to the outliers in Figure 5 above, the worst outliers here were vanadium-containing quaternary alloys for the Unary-Binary model while for the Unary-Binary-Ternary model structures containing both copper and tungsten were outliers.
Despite this, we emphasize that the binary model fundamentally captures the energy trends across the full energy range in the quaternary dataset with marginal change in the error (MAE = 0.09 eV/atom, RMSE = 0.24 eV/atom) as shown in Figure S4.
From these results, it is most likely that these potentials are interpolating the multi-element, many-body effects as opposed to an effective summation of all the binary pairwise terms, as it is well known that traditional pairwise interactions are wildly insufficient for metal alloy systems \cite{daw1993embedded,ercolessi1988simulation}.
This further shows that the diversity of data in the Unary-Binary set is sufficient to generalize beyond unary and binary systems, and that the neural network is truly "learning" atomic interactions such that it can interpolate along interactions it has not seen in training.

\begin{figure}[htpb]
  \includegraphics[width=0.95\textwidth]{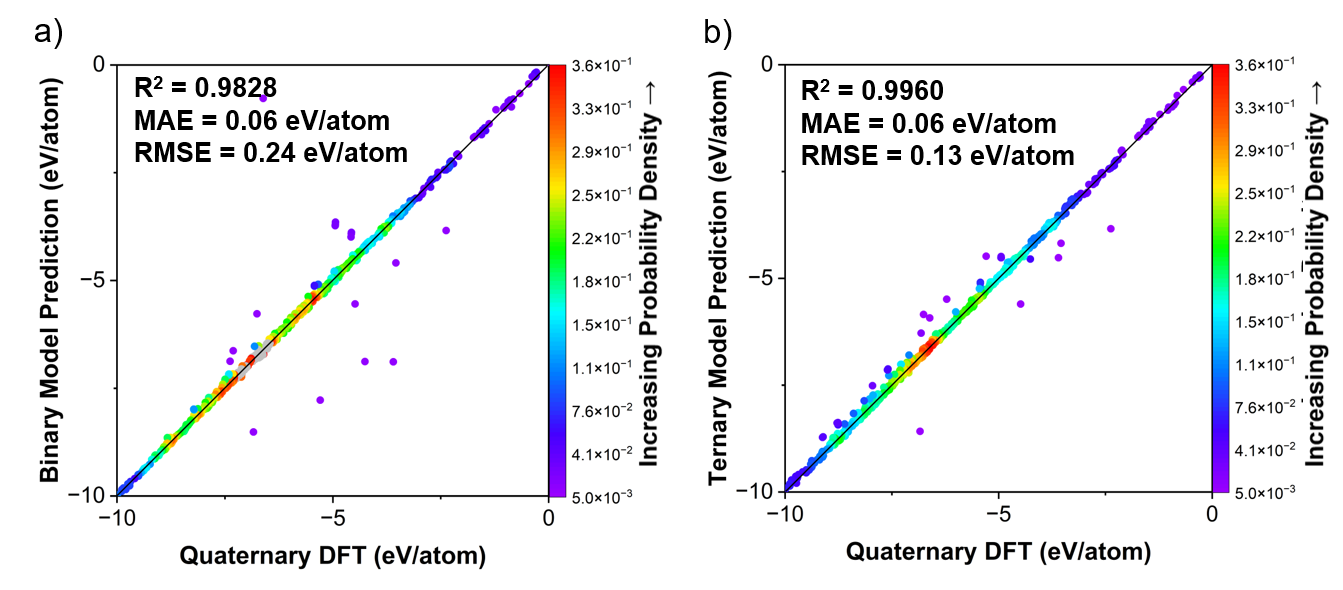}
  \caption{Parity plots comparing DFT and model predicted energy per atom for the final, hyper-parameter optimized a) Unary-Binary, and b) Unary-Binary-Ternary models on the quaternary alloy test set. Note that high energy structures are not shown. See Figure S3 for full plot.}
  \label{fig:Fg6}
\end{figure}

\subsection{Learning New Elements}

Finally, to test the ability of the model to extend to new elements and further benchmark its performance in interpolating unseen interactions from its training data, 
we supply unary-only Au data to the Unary-Binary-Ternary data.
We train an additional model using the same settings as the final hyperparameter optimized Unary-Binary-Ternary model, including the unary Au data and no Au-X interactions with the other elements. 
We then test the model on all Au-X binary systems, where each system consists Au and one of the 12 metals present in the original datasets, 
scanning a wide range of volumes and random spatial distributions of elements. 
Overall, 1170 unary Au structures were added to the original Unary-Binary-Ternary dataset, while 1,992 Au-containing binary structures were used as a test set.

Figure 7 below compares the model prediction to the DFT ground truth for low energy structures in the Au Binary test set (see Figure S5 for full plot). 
In general, we find that the model over-predicts the energy with an average error of 0.16 eV/atom but maintains a high correlation with an R${^2}$ of 0.9928. 
Moreover, the model gives consistent predictions shown by the lack of large outliers and the comparable MAE (0.21 eV/atom) and RMSE (0.26 eV/atom) errors. 
We highlight here that while the errors are higher, the good R${^2}$ value and lack of outliers shows significant accuracy in properly ranking the energy of all Au-containing binary alloys. 
This further emphasizes that the model learned how Au alloys with the other metals studied here through training on only unary Au data and non Au-containing alloys, 
capturing all energy changes from stoichiometry changes, crystal structure changes, and volume changes from this data.
This shows, in general, that to transfer learn these models to new element systems may only require small amounts of data on the specific alloy of interest, as the alchemical interactions are generally inferrable from other systems, and, presumably, new alloy data is only needed to fine tune the model to high accuracy.

\begin{figure}[htpb]
  \includegraphics[width=0.5\textwidth]{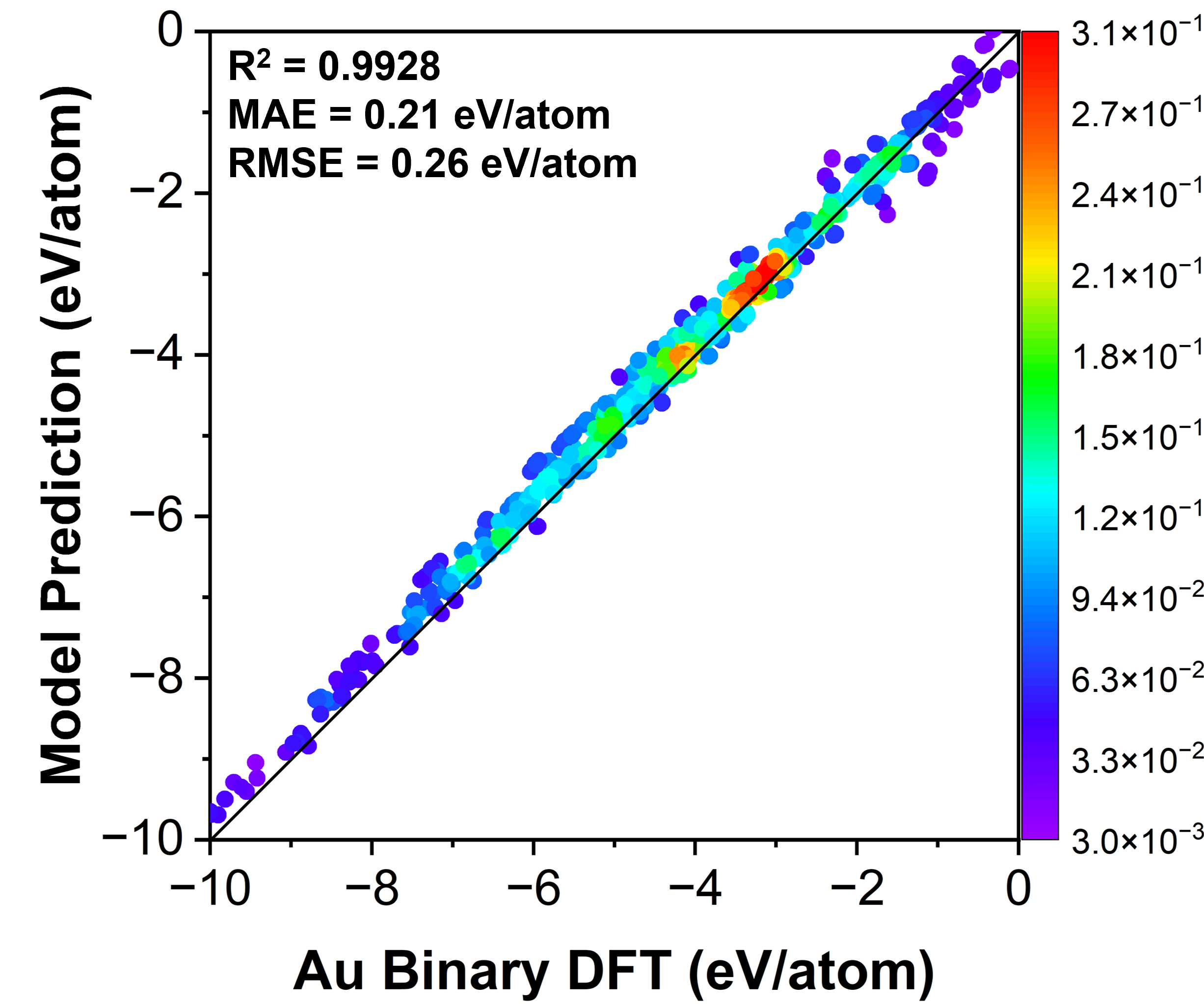}
  \caption{Parity plots comparing DFT and model predicted energy per atom for the Unary-Binary-Ternary-Au model on the binary Au binary test set.}
  \label{fig:Fg7}
\end{figure}

In addition to the above, we also explored the model errors as a function of alloying metal in the Au binary test set as shown in Figure 8 below. Prediction accuracy is broadly comparable across alloys, although Au–Al, Au–Cu, and Au–Zn exhibit lower MAEs of 0.22, 0.15, and 0.17 eV/atom, respectively. The lower errors for Cu and Zn may partly reflect electronic similarities to Au, particularly their filled d shells, and while this explanation does not directly extend to Al, these results support the model’s ability to generalize to several Au binary alloys using only unary Au and non-Au-containing training data.

\begin{figure}[htpb]
  \includegraphics[width=0.5\textwidth]{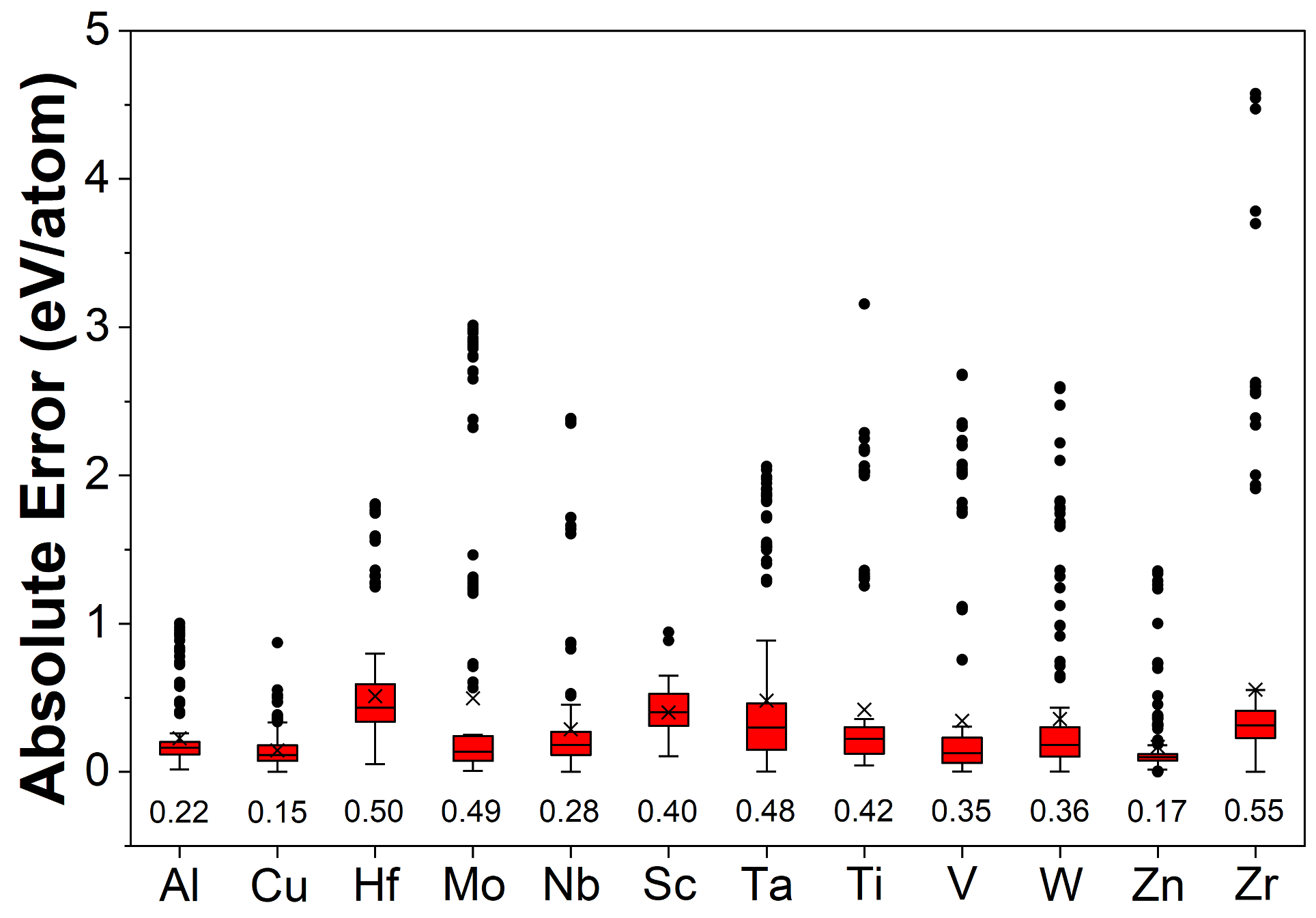}
  \caption{Box and whisker plots showing the energy prediction error of the Unary-Binary-Ternary-Au model as a function of alloying metal in the Au binary test set. Values below the box and whisker plots represent the MAE for the corresponding alloying element.}
  \label{fig:Fg8}
\end{figure}

\section{Discussion}

Overall, this work presents advances in the development of foundation-like MLIAPs for predicting and studying the properties of metals and metal alloys. 
We have utilized a data generation approach that helps to reduce data redundancy and computational costs of developing foundation-like MLIAPs while also improving generalizability of these models. 
Leveraging the exceptional interpolation capabilities of neural network based MLIAPs, HIP-NN models trained solely with 1-component and 2-component structures perform very well for n-component structures with $n \geq 3$, provided the configuration space has been sufficiently explored. 
The effectiveness of this approach has been demonstrated through accurate predictions of energies across various ternary and quaternary alloys. 

Additionally, we demonstrated that models may be trained to include previously unseen metals using only their unary data, and provide good performance on new binary alloys. 
Thus, as a result of the broad coverage of the training set, the 1-component and 2-component structures delineate an outer boundary in descriptor space, while n-component structures with  $n \geq 3$ represent interpolation points in this space. 
While previous work has primarily focused on architectural advances for developing general-purpose MLIAPs \cite{song2024general}, here we have also provided a proof-of-concept for the development of compact, yet diverse datasets for the development of general-purpose MLIAPs.

Although we have demonstrated the ability of models to  "learn between the lines" using metal and metal alloy systems with relatively simple interactions, the impact of this work goes beyond custom MLIAPs for specific chemistries. 
As we briefly noted, a major area of active research is in the development of universal/foundation MLIAPs with broad coverage of the periodic table of the elements. 
Generating a robust training dataset, even with standard GGA methods, for such universal MLIAPs is a major challenge due to the computational cost of generating the training data. 
Thus, using the approach outlined in this work, we anticipate that training universal MLIAPs, even on high-level methods, can be made significantly more data-efficient. This approach can also be extended to the creation of compact datasets for fine-tuning already existing universal MLIAPs. Future work will examine whether similar generalization can be achieved in covalent and molecular systems, where different kinds of bonding, diverse coordination environments, and bond breaking and formation events introduce additional complexity. Establishing the capabilities and limits of transferability in MLIAPs will help guide the development of compact training datasets for general-purpose MLIAPs spanning broader chemical spaces and reactive processes.

\section{Methods}

All DFT simulations for data labeling were conducted using the VASP package (version 6.6) \cite{Kresse1996PlaneWave} using the PBE functional \cite{Perdew1996GGA}.
A plane-wave cutoff of 400 eV was used with a (5,5,5) Monkhorst–Pack \cite{Monkhorst-Pack} gamma-centered K-point grid. Gaussian smearing was employed with a 0.10 eV width.
To obtain accurate forces, the energy convergence criteria was set to $1*10^{-7}$ eV.

All initial atomic configurations were built as solid crystalline alloys from the perfect crystal symmetry for the FCC, BCC, and HCP lattices,
with the lattice parameter varied over a range of volumes based on the elements in the alloy (see Table S1). Additionally, initial liquid structures were created via high temperature (3000 K) MD simulations using simple EAM potentials over a range of volumes (see Table S1).
Prior to AL sampling, random atomic position perturbations were applied as a random displacement vector chosen from a normal distribution with a standard deviation of 0.15 \AA.

Our training set generation approach was based on an active learning-based query-by-committee approach,
as has been previously utilized \cite{Grizzi_LohrALF2026,smith2021automated,kulichenko2023uncertainty,kulichenko2024data}.
As shown in Figure 1, there are 4 main steps in the active learning cycle: DFT calculations (labeling), Training Set Construction, Ensemble Training, and Uncertainty-based Dynamics (sampling).

Ensemble training is done with four distinct HIP-NN \cite{lubbers2018hierarchical} models with different randomly seeded train/validation/test splits (a 60/20/20 split is used throughout the study).
The uncertainty-based dynamics is done by running molecular dynamics on a sample structure with all four models. If the ensemble predicted standard deviation
for the energy or any force rises above a cutoff, the simulation is stopped and the uncertain frame is selected for labeling.
During these dynamics, the temperature and density are varied aggressively (see section 1 of Supporting Information) in order to induce phase transitions, melting, plasticity, and defect formation.
The new labeled structures are then added to the training set, and for every N=200 new structures added, the training sets are re-built and a new ensemble of models is trained.

Additional, extended methods, including parameter choices for the HIP-NN models and AL routine are available in the Supporting Information.

\section{Acknowledgments}

This research work was supported by the Laboratory Directed Research and Development program of Los Alamos National Laboratory. Partial funding was provided by ASC Computational Systems \& Software Environment (CSSE). AS gratefully acknowledges support from the Center for Non-Linear Studies and from the U.S. Department of Energy, Office of Basic Energy Sciences (FWP: LANLE3F2).
This research used resources provided by the Los Alamos National Laboratory (LANL) Institutional Computing Program. This work was supported by the U.S. Department of Energy (DOE) through LANL, which is operated by Triad National Security, LLC, 
for the National Nuclear Security Administration of the U.S. Department of Energy (Contract No. 89233218CNA000001). Approved for Unlimited Release LA-UR-26-28744.

\section{Competing Interests}

The Authors declare no Competing Financial or Non-Financial Interests.


\section{Data Availability}

The datasets used and/or analyzed during the current study are available from the corresponding author on reasonable request.



\bibliography{references}

@article{rupp2012fast,
  title={Fast and accurate modeling of molecular atomization energies with machine learning},
  author={Rupp, Matthias and Tkatchenko, Alexandre and M{\"u}ller, Klaus-Robert and Von Lilienfeld, O Anatole},
  journal={Physical review letters},
  volume={108},
  number={5},
  pages={058301},
  year={2012},
  publisher={APS}
}

@article{Maxson2025Materials25,
author = {Maxson, Tristan and Soyemi, Ademola and Zhang, Xinglong and Chen, Benjamin W. J. and Szilvási, Tibor},
title = {MS25: Materials Science-Focused Benchmark Data Set for Machine Learning Interatomic Potentials},
journal = {Journal of Chemical Information and Modeling},
volume = {65},
number = {15},
pages = {8097-8112},
year = {2025},
doi = {10.1021/acs.jcim.5c01262},
    note ={PMID: 40742204},
}

@article{Monkhorst-Pack,
  title = {Special points for Brillouin-zone integrations},
  author = {Monkhorst, Hendrik J. and Pack, James D.},
  journal = {Phys. Rev. B},
  volume = {13},
  issue = {12},
  pages = {5188--5192},
  numpages = {0},
  year = {1976},
  month = {Jun},
  publisher = {American Physical Society},
  doi = {10.1103/PhysRevB.13.5188},
  url = {https://link.aps.org/doi/10.1103/PhysRevB.13.5188}
}

@article{Bartok2018_Si,
  title = {Machine Learning a General-Purpose Interatomic Potential for Silicon},
  author = {Bart\'ok, Albert P. and Kermode, James and Bernstein, Noam and Cs\'anyi, G\'abor},
  journal = {Phys. Rev. X},
  volume = {8},
  issue = {4},
  pages = {041048},
  numpages = {32},
  year = {2018},
  month = {Dec},
  publisher = {American Physical Society},
  doi = {10.1103/PhysRevX.8.041048},
  url = {https://link.aps.org/doi/10.1103/PhysRevX.8.041048}
}

@article{Dicks2026,
title = {Expanding the search space of high entropy oxides and predicting synthesizability using machine learning interatomic potentials},
journal = {Computational Materials Science},
volume = {267},
pages = {114581},
year = {2026},
issn = {0927-0256},
doi = {https://doi.org/10.1016/j.commatsci.2026.114581},
url = {https://www.sciencedirect.com/science/article/pii/S092702562600100X},
author = {Oliver A. Dicks and Solveig S. Aamlid and Alannah M. Hallas and Joerg Rottler},
}

@article{Takamoto2022,
  title = {Towards universal neural network potential for material discovery applicable to arbitrary combination of 45 elements},
  author = {Takamoto, So and Shinagawa, Chikashi and Motoki, Daisuke and Nakago, Kosuke and Wenwen, Li and Kurata, Iori and Watanabe, Taku and Yayama, Yoshihiro and Iriguchi, Hiroki and Asano, Yusuke and Onodera, Tasuku and Ishii, Takafumi and Kudo, Takao and Ono, Hideki and Sawada, Ryohto and Ishitani, Ryuichiro and Ong, Marc and Yamaguchi, Taiki and Kataoka, Toshiki, Hayashi, Akihide and Charoenphakdee, Nontawat and Ibuka, Takeshi},
  journal = {Nat. Commun.},
  volume = {13},
  pages = {2991},
  year = {2022},
}

@article{Chen_Ong2022,
  title = {A universal graph deep learning interatomic potential for the periodic table },
  author = {Chen, Chi and Ong, Shyue P.},
  journal = {Nat. Comput. Sci.},
  volume = {2},
  pages = {718-728},
  year = {2022},
}

@article{Sivak2025,
  title = {Discovering High-Entropy Oxides with a Machine-Learning Interatomic Potential},
  author = {Sivak, Jacob T. and Almishal, Saeed S. I. and Caucci, Mary Kathleen and Tan, Yueze and Srikanth, Dhiya and Petruska, Joseph and Furst, Matthew and Chen, Long-Qing and Rost, Christina M. and Maria, Jon-Paul and Sinnott, Susan B.},
  journal = {Phys. Rev. Lett.},
  volume = {134},
  issue = {21},
  pages = {216101},
  numpages = {6},
  year = {2025},
  month = {May},
  publisher = {American Physical Society},
  doi = {10.1103/PhysRevLett.134.216101},
  url = {https://link.aps.org/doi/10.1103/PhysRevLett.134.216101}
}

@article{Csanyi2025,
  title = {Discovering High-Entropy Oxides with a Machine-Learning Interatomic Potential},
  author = {Batatia, Ilyes and Batzner, Simon and Kovács, Dávid Péter and Musaelian, Albert and Simm, Gregor N. C. and Drautz, Ralf and Ortner, Christoph and Kozinsky, Boris and Csányi, Gábor},
  journal = {Nat. Mech. Intell.},
  volume = {7},
  pages = {56-67},
  year = {2025},
  month = {January},
  doi = {10.1038/s42256-024-00956-x},
}

@article{Tateishi_PFP_2026,
    author = {Tateishi, Yusuke and Botha, Louise M. and Kozhukhova, Alina E. and Sugimoto, Manabu and Aika, Ken-ichi and Bessarabov, Dmitri G.},
    title = {Molecular dynamics study of catalytic H2/O2 recombination on Pd, Pt, Cu, Ag, and Au nanoclusters using the universal neural network potential},
    journal = {Chemical Science},
    volume = {17},
    number = {11},
    pages = {5542-5553},
    year = {2026},
    month = {03},
    issn = {2041-6520},
    doi = {10.1039/d5sc04712a},
    url = {https://doi.org/10.1039/d5sc04712a},
    eprint = {https://pubs.rsc.org/sc/article-pdf/17/11/5542/10664513/d5sc04712a.pdf},
}

@article{Rowe_Deringer2020_Carbon,
    author = {Rowe, Patrick and Deringer, Volker L. and Gasparotto, Piero and Csányi, Gábor and Michaelides, Angelos},
    title = {An accurate and transferable machine learning potential for carbon},
    journal = {The Journal of Chemical Physics},
    volume = {153},
    number = {3},
    pages = {034702},
    year = {2020},
    month = {07},
    issn = {0021-9606},
    doi = {10.1063/5.0005084},
    url = {https://doi.org/10.1063/5.0005084},
    eprint = {https://pubs.aip.org/aip/jcp/article-pdf/doi/10.1063/5.0005084/13464163/034702_1_online.pdf},
}

@article{Ugwumadu2023,
title = {Simulation of multi-shell fullerenes using Machine-Learning Gaussian Approximation Potential},
journal = {Carbon Trends},
volume = {10},
pages = {100239},
year = {2023},
issn = {2667-0569},
doi = {https://doi.org/10.1016/j.cartre.2022.100239},
url = {https://www.sciencedirect.com/science/article/pii/S2667056922000955},
author = {C. Ugwumadu and K. Nepal and R. Thapa and Y.G. Lee and Y. {Al Majali} and J. Trembly and D.A. Drabold},
}

@article{Maxson2024Practices,
author = {Maxson, Tristan and Soyemi, Ademola and Chen, Benjamin W. J. and Szilvási, Tibor},
title = {Enhancing the Quality and Reliability of Machine Learning Interatomic Potentials through Better Reporting Practices},
journal = {The Journal of Physical Chemistry C},
volume = {128},
number = {16},
pages = {6524-6537},
year = {2024},
doi = {10.1021/acs.jpcc.4c00028},
}

@article{Maxson2025AgNP,
author = {Maxson, Tristan and Szilvási, Tibor},
title = {Metal-Support Interactions Reshape Nanoparticle Catalyst Surfaces},
journal = {Angewandte Chemie Novit},
volume = {1},
number = {1},
pages = {e70008},
doi = {https://doi.org/10.1002/anov.70008},
url = {https://onlinelibrary.wiley.com/doi/abs/10.1002/anov.70008},
eprint = {https://onlinelibrary.wiley.com/doi/pdf/10.1002/anov.70008},
year = {2025}
}

@article{Soyemi2025Na2So4,
author = {Soyemi, Ademola and Szilvási, Tibor},
title = {Modeling the Behavior of Complex Aqueous Electrolytes Using Machine Learning Interatomic Potentials: The Case of Sodium Sulfate},
journal = {The Journal of Physical Chemistry B},
volume = {129},
number = {37},
pages = {9405-9417},
year = {2025},
doi = {10.1021/acs.jpcb.5c02306},
    note ={PMID: 40685593},
}

@article{Kim_Leveraging2026,
author = {Kim, So Yeon and Park, Yang Jeong and Li, Ju},
title = {Leveraging neural network interatomic potentials for a foundation model of chemistry},
journal = {npj Computational Materails},
year = {2026},
doi = {https://doi.org/10.1038/s41524-026-02167-x},
}

@article{Soyemi2025Interface,
    author = {Soyemi, Ademola and Szilvási, Tibor},
    title = {Cation dominated but negatively charged Na2SO4,aq–graphene interfaces},
    journal = {The Journal of Chemical Physics},
    volume = {164},
    number = {9},
    pages = {094702},
    year = {2026},
    month = {03},
    issn = {0021-9606},
    doi = {10.1063/5.0309415},
    url = {https://doi.org/10.1063/5.0309415},
    eprint = {https://pubs.aip.org/aip/jcp/article-pdf/doi/10.1063/5.0309415/20925115/094702_1_5.0309415.pdf},
}

@article{desai2021parsimonious,
  title={Parsimonious neural networks learn interpretable physical laws},
  author={Desai, Saaketh and Strachan, Alejandro},
  journal={Scientific reports},
  volume={11},
  number={1},
  pages={12761},
  year={2021},
  publisher={Nature Publishing Group UK London}
}

@article{butler2018machine,
  title={Machine learning for molecular and materials science},
  author={Butler, Keith T and Davies, Daniel W and Cartwright, Hugh and Isayev, Olexandr and Walsh, Aron},
  journal={Nature},
  volume={559},
  number={7715},
  pages={547--555},
  year={2018},
  publisher={Nature Publishing Group UK London}
}

@article{ramprasad2017machine,
  title={Machine learning in materials informatics: recent applications and prospects},
  author={Ramprasad, Rampi and Batra, Rohit and Pilania, Ghanshyam and Mannodi-Kanakkithodi, Arun and Kim, Chiho},
  journal={npj Computational Materials},
  volume={3},
  number={1},
  pages={54},
  year={2017},
  publisher={Nature Publishing Group UK London}
}

@article{Grizzi_LohrALF2026,
  title={ALF: Open-Source Active Learning Framework for Atomistic Modeling},
  author={Grizzi, Vitor F. and Lohr, Patrick J. and Fedik, Nikita and Anaya, Steven and Averitt, Jared and Gomes, Joseph and Li, Ying Wai and Barros, Kipton and Messerly, Richard A. and Janssen, Jan and Kulichenko, Maksim and Smith, Justin and Tretiak, Sergei and Nebgen, Benjamin T.},
  journal={Journal of Chemical Theory and Computation},
  year={2026},
}

@article{smith2019approaching,
  title={Approaching coupled cluster accuracy with a general-purpose neural network potential through transfer learning},
  author={Smith, Justin S and Nebgen, Benjamin T and Zubatyuk, Roman and Lubbers, Nicholas and Devereux, Christian and Barros, Kipton and Tretiak, Sergei and Isayev, Olexandr and Roitberg, Adrian E},
  journal={Nature communications},
  volume={10},
  year={2019},
}

@article{zuo2020performance,
  title={Performance and cost assessment of machine learning interatomic potentials},
  author={Zuo, Yunxing and Chen, Chi and Li, Xiangguo and Deng, Zhi and Chen, Yiming and Behler, Jörg and Cs{\'a}nyi, G{\'a}bor and Shapeev, Alexander V and Thompson, Aidan P and Wood, Mitchell A and others},
  journal={The Journal of Physical Chemistry A},
  volume={124},
  number={4},
  pages={731--745},
  year={2020},
  publisher={ACS Publications}
}

@article{smith2017ani,
  title={ANI-1: an extensible neural network potential with DFT accuracy at force field computational cost},
  author={Smith, Justin S and Isayev, Olexandr and Roitberg, Adrian E},
  journal={Chemical science},
  volume={8},
  number={4},
  pages={3192--3203},
  year={2017},
  publisher={Royal Society of Chemistry}
}

@article{chmiela2018towards,
  title={Towards exact molecular dynamics simulations with machine-learned force fields},
  author={Chmiela, Stefan and Sauceda, Huziel E and M{\"u}ller, Klaus-Robert and Tkatchenko, Alexandre},
  journal={Nature communications},
  volume={9},
  number={1},
  pages={3887},
  year={2018},
  publisher={Nature Publishing Group UK London}
}

@article{smith2018less,
  title={Less is more: Sampling chemical space with active learning},
  author={Smith, Justin S and Nebgen, Ben and Lubbers, Nicholas and Isayev, Olexandr and Roitberg, Adrian E},
  journal={The Journal of chemical physics},
  volume={148},
  number={24},
  year={2018},
  publisher={AIP Publishing}
}

@article{smith2021automated,
  title={Automated discovery of a robust interatomic potential for aluminum},
  author={Smith, Justin S and Nebgen, Benjamin and Mathew, Nithin and Chen, Jie and Lubbers, Nicholas and Burakovsky, Leonid and Tretiak, Sergei and Nam, Hai Ah and Germann, Timothy and Fensin, Saryu and others},
  journal={Nature communications},
  volume={12},
  number={1},
  pages={1257},
  year={2021},
  publisher={Nature Publishing Group UK London}
}

@article{kulichenko2024data,
  title={Data generation for machine learning interatomic potentials and beyond},
  author={Kulichenko, Maksim and Nebgen, Benjamin and Lubbers, Nicholas and Smith, Justin S and Barros, Kipton and Allen, Alice EA and Habib, Adela and Shinkle, Emily and Fedik, Nikita and Li, Ying Wai and others},
  journal={Chemical Reviews},
  volume={124},
  number={24},
  pages={13681--13714},
  year={2024},
  publisher={ACS Publications}
}

@article{lindsey2025chimes,
  title={ChIMES Carbon 2.0: A transferable machine-learned interatomic model harnessing multifidelity training data},
  author={Lindsey, Rebecca K and Bastea, Sorin and Hamel, Sebastien and Lyu, Yanjun and Goldman, Nir and Lordi, Vincenzo},
  journal={npj Computational Materials},
  volume={11},
  number={1},
  pages={26},
  year={2025},
  publisher={Nature Publishing Group UK London}
}

@article{vandermause2020fly,
  title={On-the-fly active learning of interpretable Bayesian force fields for atomistic rare events},
  author={Vandermause, Jonathan and Torrisi, Steven B and Batzner, Simon and Xie, Yu and Sun, Lixin and Kolpak, Alexie M and Kozinsky, Boris},
  journal={npj Computational Materials},
  volume={6},
  number={1},
  pages={20},
  year={2020},
  publisher={Nature Publishing Group UK London}
}

@article{xie2021bayesian,
  title={Bayesian force fields from active learning for simulation of inter-dimensional transformation of stanene},
  author={Xie, Yu and Vandermause, Jonathan and Sun, Lixin and Cepellotti, Andrea and Kozinsky, Boris},
  journal={npj Computational Materials},
  volume={7},
  number={1},
  pages={40},
  year={2021},
  publisher={Nature Publishing Group UK London}
}

@article{kovacs2025mace,
  title={Mace-off: Short-range transferable machine learning force fields for organic molecules},
  author={Kov{\'a}cs, D{\'a}vid P{\'e}ter and Moore, J Harry and Browning, Nicholas J and Batatia, Ilyes and Horton, Joshua T and Pu, Yixuan and Kapil, Venkat and Witt, William C and Magdau, Ioan-Bogdan and Cole, Daniel J and others},
  journal={Journal of the American Chemical Society},
  volume={147},
  number={21},
  pages={17598--17611},
  year={2025},
  publisher={ACS Publications}
}

@article{allen2024learning,
  title={Learning together: Towards foundation models for machine learning interatomic potentials with meta-learning},
  author={Allen, Alice EA and Lubbers, Nicholas and Matin, Sakib and Smith, Justin and Messerly, Richard and Tretiak, Sergei and Barros, Kipton},
  journal={npj computational materials},
  volume={10},
  number={1},
  pages={154},
  year={2024},
  publisher={Nature Publishing Group UK London}
}

@article{zhang2024exploring,
  title={Exploring the frontiers of condensed-phase chemistry with a general reactive machine learning potential},
  author={Zhang, Shuhao and Mako{\'s}, Ma{\l}gorzata Z and Jadrich, Ryan B and Kraka, Elfi and Barros, Kipton and Nebgen, Benjamin T and Tretiak, Sergei and Isayev, Olexandr and Lubbers, Nicholas and Messerly, Richard A and others},
  journal={Nature Chemistry},
  volume={16},
  number={5},
  pages={727--734},
  year={2024},
  publisher={Nature Publishing Group UK London}
}

@article{f2025improving,
  title={Improving bond dissociations of reactive machine learning potentials through physics-constrained data augmentation},
  author={F. dos Santos, Luan G and Nebgen, Benjamin T and Allen, Alice EA and Hamilton, Brenden W and Matin, Sakib and Smith, Justin S and Messerly, Richard A},
  journal={Journal of Chemical Information and Modeling},
  volume={65},
  number={3},
  pages={1198--1210},
  year={2025},
  publisher={ACS Publications}
}

@article{sahrmann2026learning,
  title={Learning Thermal Response Forces: A Method for Extending the Thermodynamic Transferability of Coarse-Grained Models via Machine-Learning},
  author={Sahrmann, Patrick G and Nebgen, Benjamin T and Barros, Kipton and Hamilton, Brenden W},
  journal={The Journal of Physical Chemistry B},
  volume={130},
  number={33},
  pages={8442--8448},
  year={2026},
  publisher={ACS Publications}
}

@article{sahrmann2026bridging,
  title={Bridging Molecular Scales with Implicit Score Matching for Bottom-Up Coarse Graining},
  author={Sahrmann, Patrick G and Lubbers, Nicholas and Nebgen, Benjamin T and Shinkle, Emily and Tretiak, Sergei and Barros, Kipton and Hamilton, Brenden W},
  journal={arXiv preprint arXiv:2609.32040},
  year={2026}
}

@article{hamilton2026stoichiometry,
  title={Stoichiometry dependent properties of cerium hydride: An active learning developed interatomic potential study},
  author={Hamilton, Brenden W and Jones, Travis E and Germann, Timothy C and Nebgen, Benjamin T},
  journal={Physical Review Materials},
  volume={10},
  number={7},
  pages={073603},
  year={2026},
  publisher={APS}
}

@article{huan2017universal,
  title={A universal strategy for the creation of machine learning-based atomistic force fields},
  author={Huan, Tran Doan and Batra, Rohit and Chapman, James and Krishnan, Sridevi and Chen, Lihua and Ramprasad, Rampi},
  journal={npj Computational Materials},
  volume={3},
  number={1},
  pages={37},
  year={2017},
  publisher={Nature Publishing Group UK London}
}

@article{niblett2025transferability,
  title={Transferability of Data Sets between Machine-Learned Interatomic Potential Algorithms},
  author={Niblett, Samuel P and Kourtis, Panagiotis and Magdau, Ioan-Bogdan and Grey, Clare P and Cs{\'a}nyi, G{\'a}bor},
  journal={Journal of Chemical Theory and Computation},
  year={2025},
  publisher={ACS Publications}
}

@article{montes2022training,
  title={Training data selection for accuracy and transferability of interatomic potentials},
  author={Montes de Oca Zapiain, David and Wood, Mitchell A and Lubbers, Nicholas and Pereyra, Carlos Z and Thompson, Aidan P and Perez, Danny},
  journal={npj Computational Materials},
  volume={8},
  number={1},
  pages={189},
  year={2022},
  publisher={Nature Publishing Group UK London}
}

@article{schutt2017quantum,
  title={Quantum-chemical insights from deep tensor neural networks},
  author={Sch{\"u}tt, Kristof T and Arbabzadah, Farhad and Chmiela, Stefan and M{\"u}ller, Klaus R and Tkatchenko, Alexandre},
  journal={Nature communications},
  volume={8},
  number={1},
  pages={13890},
  year={2017},
  publisher={Nature Publishing Group UK London}
}

@article{bartok2010gaussian,
  title={Gaussian approximation potentials: The accuracy of quantum mechanics, without the electrons},
  author={Bart{\'o}k, Albert P and Payne, Mike C and Kondor, Risi and Cs{\'a}nyi, G{\'a}bor},
  journal={Physical review letters},
  volume={104},
  number={13},
  pages={136403},
  year={2010},
  publisher={APS}
}

@article{keith2021combining,
  title={Combining machine learning and computational chemistry for predictive insights into chemical systems},
  author={Keith, John A and Vassilev-Galindo, Valentin and Cheng, Bingqing and Chmiela, Stefan and Gastegger, Michael and Muller, Klaus-Robert and Tkatchenko, Alexandre},
  journal={Chemical reviews},
  volume={121},
  number={16},
  pages={9816--9872},
  year={2021},
  publisher={ACS Publications}
}

@article{daw1993embedded,
  title={The embedded-atom method: a review of theory and applications},
  author={Daw, Murray S and Foiles, Stephen M and Baskes, Michael I},
  journal={Materials Science Reports},
  volume={9},
  number={7-8},
  pages={251--310},
  year={1993},
  publisher={Elsevier}
}

@article{ercolessi1988simulation,
  title={Simulation of gold in the glue model},
  author={Ercolessi, Furio and Parrinello, M and Tosatti, E},
  journal={Philosophical magazine A},
  volume={58},
  number={1},
  pages={213--226},
  year={1988},
  publisher={Taylor \& Francis}
}

@article{song2024general,
  title={General-purpose machine-learned potential for 16 elemental metals and their alloys},
  author={Song, Keke and Zhao, Rui and Liu, Jiahui and Wang, Yanzhou and Lindgren, Eric and Wang, Yong and Chen, Shunda and Xu, Ke and Liang, Ting and Ying, Penghua and others},
  journal={Nature Communications},
  volume={15},
  number={1},
  pages={10208},
  year={2024},
  publisher={Nature Publishing Group UK London}
}

@article{lubbers2018hierarchical,
  title={Hierarchical modeling of molecular energies using a deep neural network},
  author={Lubbers, Nicholas and Smith, Justin S and Barros, Kipton},
  journal={The Journal of chemical physics},
  volume={148},
  number={24},
  year={2018},
  publisher={AIP Publishing}
}

@article{kulichenko2023uncertainty,
  title={Uncertainty-driven dynamics for active learning of interatomic potentials},
  author={Kulichenko, Maksim and Barros, Kipton and Lubbers, Nicholas and Li, Ying Wai and Messerly, Richard and Tretiak, Sergei and Smith, Justin S and Nebgen, Benjamin},
  journal={Nature computational science},
  volume={3},
  number={3},
  pages={230--239},
  year={2023},
  publisher={Nature Publishing Group US New York}
}

@article{batatia2025foundation,
  title={A foundation model for atomistic materials chemistry},
  author={Batatia, Ilyes and Benner, Philipp and Chiang, Yuan and Elena, Alin M and Kov{\'a}cs, D{\'a}vid P and Riebesell, Janosh and Advincula, Xavier R and Asta, Mark and Avaylon, Matthew and Baldwin, William J and others},
  journal={The Journal of chemical physics},
  volume={163},
  number={18},
  year={2025},
  publisher={AIP Publishing}
}

@article{hamilton2023using,
  title = {Using limited neural networks to assess relative mechanistic influence on shock heating in granular solids},
  author = {Hamilton, Brenden W. and Germann, Timothy C.},
  journal = {Phys. Rev. Mater.},
  volume = {7},
  issue = {8},
  pages = {085601},
  numpages = {8},
  year = {2023},
  month = {Aug},
  publisher = {American Physical Society},
  doi = {10.1103/PhysRevMaterials.7.085601},
  url = {https://link.aps.org/doi/10.1103/PhysRevMaterials.7.085601}
}

@article{hamilton2023high,
  title={High-pressure and temperature neural network reactive force field for energetic materials},
  author={Hamilton, Brenden W and Yoo, Pilsun and Sakano, Michael N and Islam, Md Mahbubul and Strachan, Alejandro},
  journal={The Journal of Chemical Physics},
  volume={158},
  number={14},
  year={2023},
  pages={144117},
  publisher={AIP Publishing}
}

@article{Perdew1996GGA,
author = {Perdew, John P. and Burke, Kieron and Ernzerhof, Matthias},
doi = {10.1103/PhysRevLett.77.3865},
issn = {10797114},
journal = {Physical Review Letters},
number = {18},
pages = {3865--3868},
pmid = {10062328},
title = {{Generalized gradient approximation made simple}},
volume = {77},
year = {1996}
}

@article{Kresse1996PlaneWave,
author = {Kresse, G. and Furthmuller, J},
doi = {10.1021/acs.jpca.0c01375},
issn = {15205215},
journal = {Physical Review B},
number = {16},
pages = {11169--11186},
pmid = {32338905},
title = {{Efficient iterative schemes for ab initio total-energy calculations using a plane-wave basis set}},
volume = {54},
year = {1996}
}

@article{Yoo2021NNRF,
author = {Yoo, Pilsun and Sakano, Michael and Desai, Saaketh and Islam, Md Mahbubul and Liao, Peilin and Strachan, Alejandro},
doi = {10.1038/s41524-020-00484-3},
issn = {2057-3960},
journal = {npj Computational Materials},
number = {1},
pages = {1--10},
publisher = {Springer US},
title = {{Neural network reactive force field for C, H, N, and O systems}},
url = {http://dx.doi.org/10.1038/s41524-020-00484-3},
volume = {7},
year = {2021}
}

\end{document}